\documentclass[12pt,a4paper,twoside]{article}
\usepackage{epsfig}
\usepackage{baltlat1}
\usepackage{wrapfig}
\begin{document}
\ \
\vspace{0.5mm}
\hyphenation{e-xam-ple co-llap-se s-phe-ri-cal}

\setcounter{page}{1}
\vspace{8mm}

\titlehead{Baltic Astronomy, vol.~12, XXX--XXX, 2003.}

\titleb{Modelling of the continuum emission from Class\,0 sources}

\begin{authorl}
\authorb{M.~Rengel}{1},
\authorb{D.~Froebrich}{2,1},
\authorb{S.~Wolf}{3} and
\authorb{J.~Eisl\"{o}ffel}{1}
\end{authorl}

\begin{addressl}
\addressb{1}{Th\"uringer Landessternwarte Tautenburg,
Sternwarte 5, 07778 Tautenburg, Germany}

\addressb{2}{Dublin Institute for Advanced Studies,
5 Merrion Square, Dublin 2, Ireland}

\addressb{3}{California Institute of Technology, 1201 East California Blvd.,
105-24, Pasadena, CA91125, USA}
\end{addressl}

\submitb{Received October 15, 2003}

\begin{abstract}
Class\,0 sources are objects representing the earliest phase of
the protostellar evolution. Since they are highly obscured by an
extended dusty envelope, these objects emit mainly in the
far-infrared to millimetre wavelength range. The analysis of their
spectral energy distributions with wide wavelength coverage allows
to determine the bolometric temperature and luminosity. However, a
more detailed physical interpretation of the internal physical
structure of these objects requires radiative transfer modelling.
We present modelling results of spectral energy distributions of a
sample of nine Class\,0 sources in the Perseus and Orion molecular
clouds. The SEDs have been simulated using a radiative transfer
code based on the Monte Carlo method. We find that a spherically
symmetric model for the youngest Class\,0 sources allows to
reproduce the observed SEDs reasonably well. From our modelling we
derive physical parameters of our sources, such as their mass,
density distribution, size, etc. We find a density structure of
$\rho \sim r^{-2}$ for the collapsing cores at young ages,
evolving to $\rho \sim r^{-3/2}$ at later times.
\end{abstract}


\resthead{Modelling of the continuum emission from Class\,0 sources}{M. Rengel,
D. Froebrich, S. Wolf, J. Eisl\"{o}ffel}



\sectionb{1}{Introduction}

A Class\,0 source is an object representing the earliest phase of the protostellar evolution
(e.g. Andr\'e et al. 1993). It consists of a central protostellar
object surrounded by an infalling dusty envelope and a flattened
accretion disk. Many (presumably all) Class 0 sources are
associated with a bipolar molecular outflow. Because of their
difficult detection (their photospheres are highly obscured and
the objects spend only a short time ($\sim$ $10^{4}$ yr) in this evolutionary phase),
and the lack of knowledge in model protostellar
parameters (distribution, composition and size of dust, density
and temperature distribution, etc.), constraining physical
properties to Class\,0 objects is quite difficult. Continuum
sub-mm observations of Class\,0 sources, however, let us detect
dust emission of the massive circumstellar envelope of these
sources. Information from the dust emission, a physical model of
the extended envelope as a pre-requisite, and the implementation of
techniques like the blackbody and the envelope fitting procedures,
open a way to interpret the structure of Class\,0 sources.

\sectionb{2}{An envelope model}


How can we describe the protostellar emission from the envelope? We adopt the
standard envelope model (Adams 1991) to keep the problem as simple as possible.
The circular symmetry of the observed emission (Rengel et al. 2004a), and the
lack of significant internal structure justifies the simplicity of a spherical
model case. If the emission is optically thin, the observed intensity for a
spherical symmetric protostellar envelope at an impact parameter b is given by
Eq.\,1 assuming single power-laws of the opacity, the radial temperature and
density distribution ($\beta$, q and p, respectively).

\begin{equation}\label{i}
I_{\nu}(b)=2\kappa_{\nu}\int^{r_{0}}_{b}
B_{\nu}\left[T_{d}(r)\right]\rho(r)\frac{r}{\sqrt{r^{2}-b^{2}}}dr
\end{equation}

$r_{0}$ is the outer radius,
$\rho$ is the density, $T_{d}$ the dust temperature,
$\kappa_{\nu}$ the opacity of the dust grains and
$B_{\nu}[T_{d}(r)]$ the Planck function at dust temperature T$_d$. If
the emission is in the Rayleigh-Jeans limit and if $r_{0}$$\gg$ b,
Eq.\,1 can be approximated to
$I_{\nu}(b)/I_{\nu}(0)=(b/b_{0})^{-m}$ (where m is the power-law index of the
observed intensity profile, and b$_{0}$ is the normalization factor to the peak emission).

Results from the standard theory provide the first direct insights into
observable estimations. Nevertheless, the temperature profile will diverge from
a single power-law q as the envelope becomes optically thick at the primary
wavelengths of energy transport (inner portion of the envelope) (e.g. Shirley,
Evans \& Rawlings 2002). Here it becomes necessary to calculate the temperature
distribution self-consistently by implementation of a radiative transfer code.
To calculate the observables (temperature and density distributions, and
intensity map) we implemented a 1D spherical symmetric radiative transfer code
(MC3D, Wolf et al. 1999). The model consists  of a central heating source, with
stellar luminosity $L_{*}$ and an effective temperature $T_{*}$, embedded
within a spherical symmetric dust envelope of mass $M_{env}$. The envelope is
specified by an outer radius $R_{out}$, a density power-law index p, and inner
radius $r_{in}$, which is set by the dust destruction radius $r_{sub}$, adopted
to a temperature of 1500\,K.

\begin{figure}[t]
\centering
\includegraphics[width=5.2cm, bb=15 180 600 710]{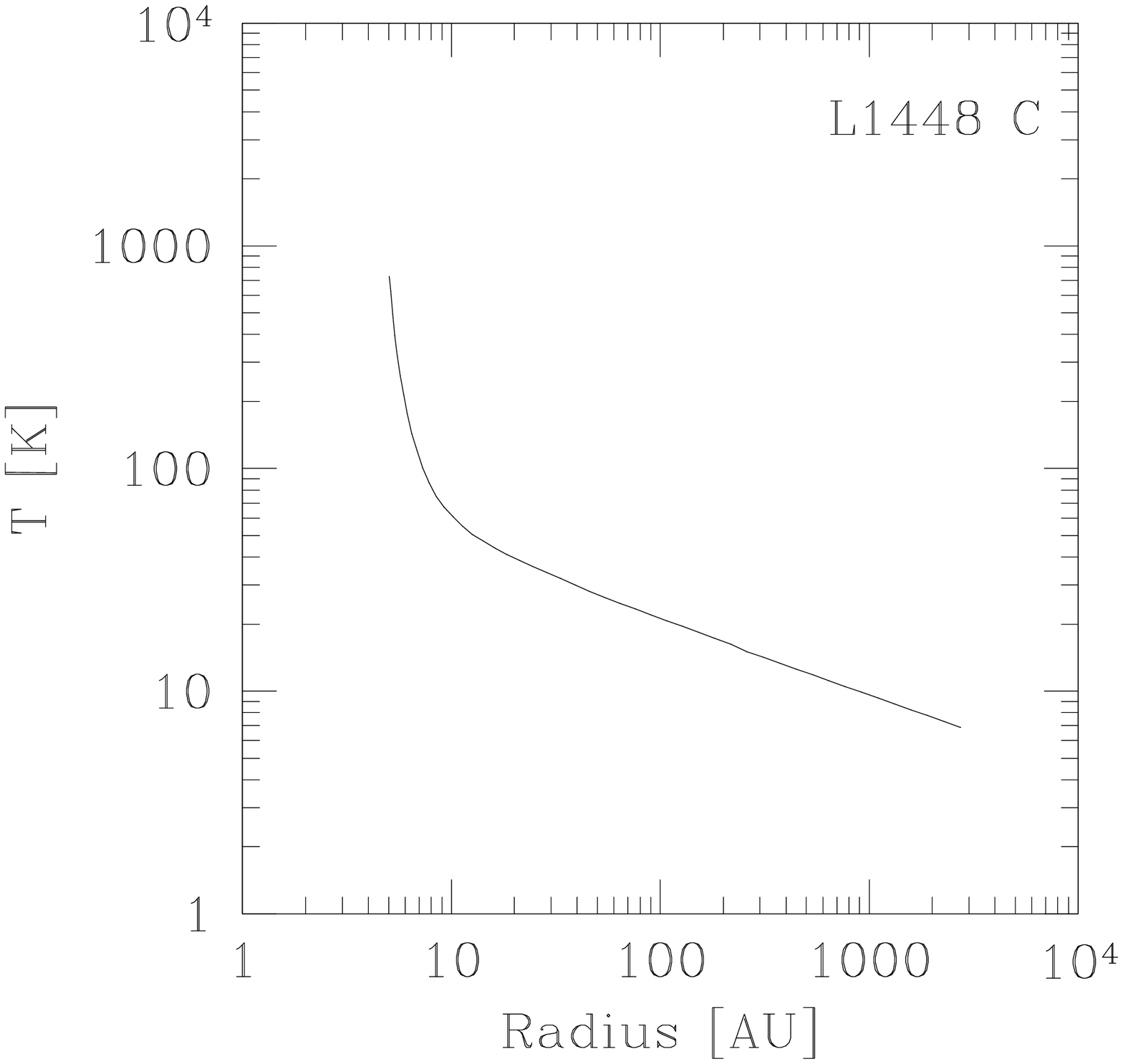}\hspace{0.2cm}
\includegraphics[width=5.0cm, bb=15 150 600 710]{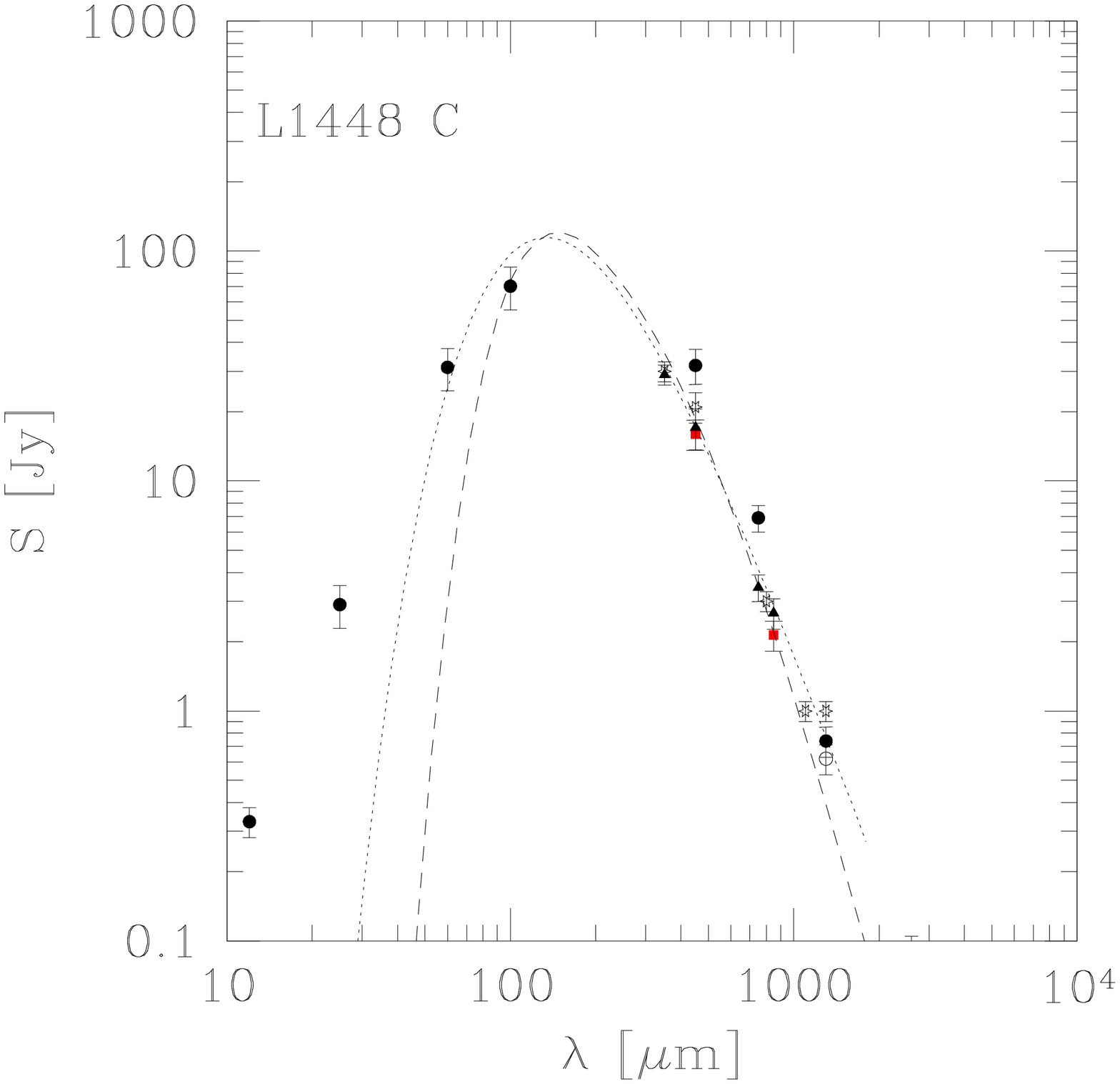}
\vspace{-0.3cm}
\caption{\label{mod1} Temperature profile (left) and best
simulated and observed SEDs (right, short dashed and dotted lines,
respectively). Details regarding the summarized target list at several
wavelengths from the literature are given in Rengel et al.
(2004b).}
\end{figure}

\sectionb{3}{Modeling continuum submillimetre emission }

First constrains to physical properties (spectral indices, masses, radial
profiles and sizes) of 15 objects are derived from continuum SCUBA imaging of
six star formation regions in the Perseus and Orion molecular cloud complexes
(Rengel et al. 2004a). Accurate SEDs are computed for nine sources combining
existing multi-wavelength surveys with new sub-mm data (Rengel et al. 2004b),
and fitting a modified blackbody curve to the fluxes (Froebrich et al. 2003).
We derived the bolometric temperature and luminosity, size of the envelope and
sub-mm slope for each source. Furthermore, dust temperature distribution, SED,
and an intensity map are derived self-consistently by the MC3D code. Physical
parameters of the sources (e.g. envelope masses, density distributions, sizes
and sublimation radius) were derived by finding the consistency between
observed and modelled SEDs.


We created a bolometric luminosity-temperature diagram for the nine
Class\,0 sources and compare the positions of the objects with an evolutionary
model in order to derive the age and mass of the envelope of each source.

\sectionb{4}{Results}
 
In Fig.\,1, the calculated temperature profile (left), and the observed
and best simulated SEDs for L1448\,C are shown as an example. Table\,1 lists
the characteristics of the best-fitting model parameters.
The envelope emission is calculated as an image from the MC3D code, and
convolved by the SCUBA beam. The simulated intensity profile is calculated at
850 $\mu$m and compared with the observed profile to test the symmetric model.


Is there a correlation of the power-law index p with time? We plot the
estimated values of p as a function of t. Fig.\,2 suggests a density structure
of $\rho \propto r^{-2}$ at younger ages, evolving to $\rho \propto r^{-3/2}$
at later times. In order to investigate this with a larger spread, a Class\,1
sample is mandatory.

\sectionb{5}{Conclusions}


\begin{figure}[t]
\parbox{5.5cm}{\includegraphics[height=6.5cm, bb=15 20 600 750,
angle=-90]{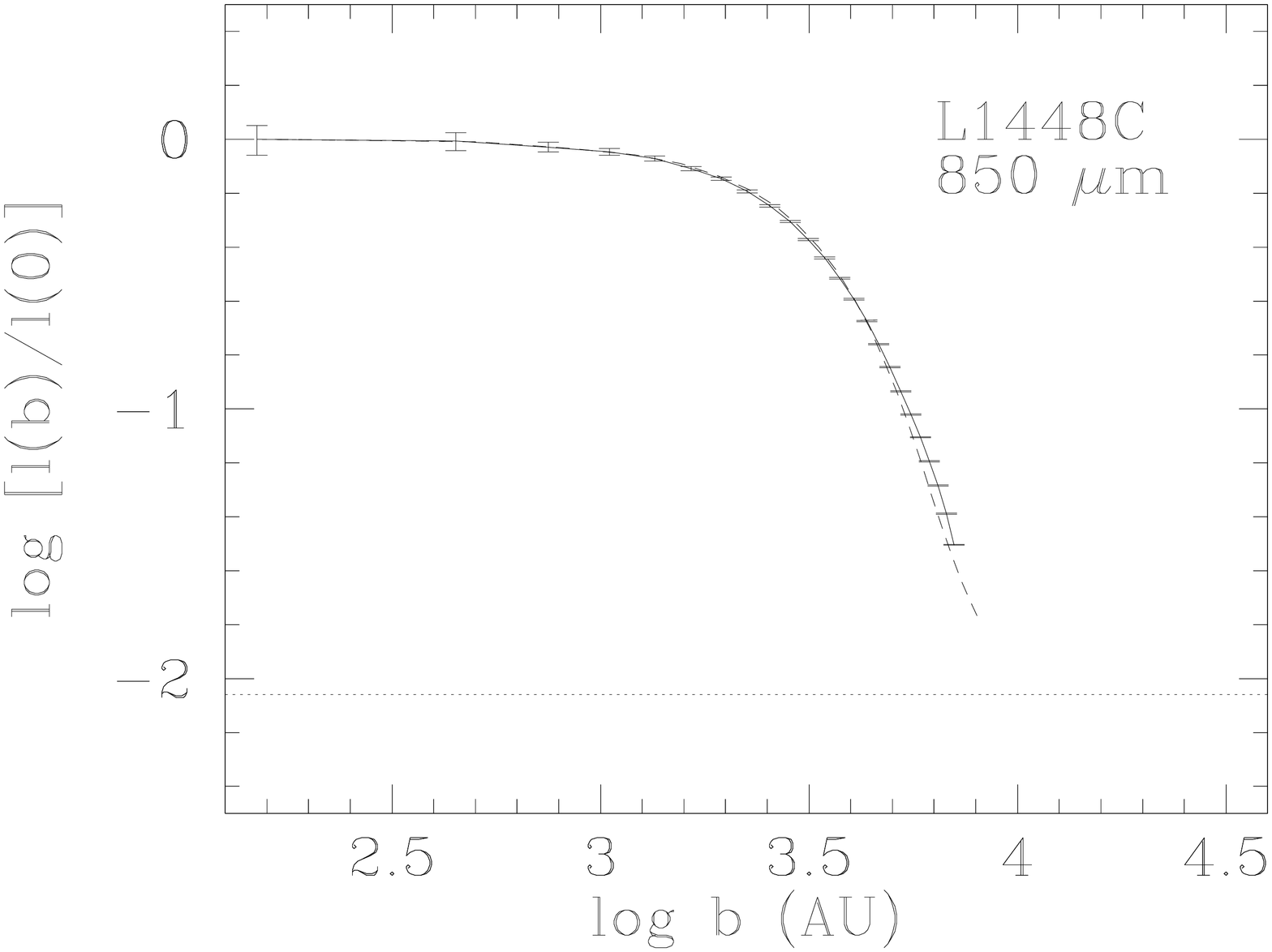}} \hfill
\parbox{5.5cm}{\includegraphics[height=5.0cm, bb=0 150 590 720]{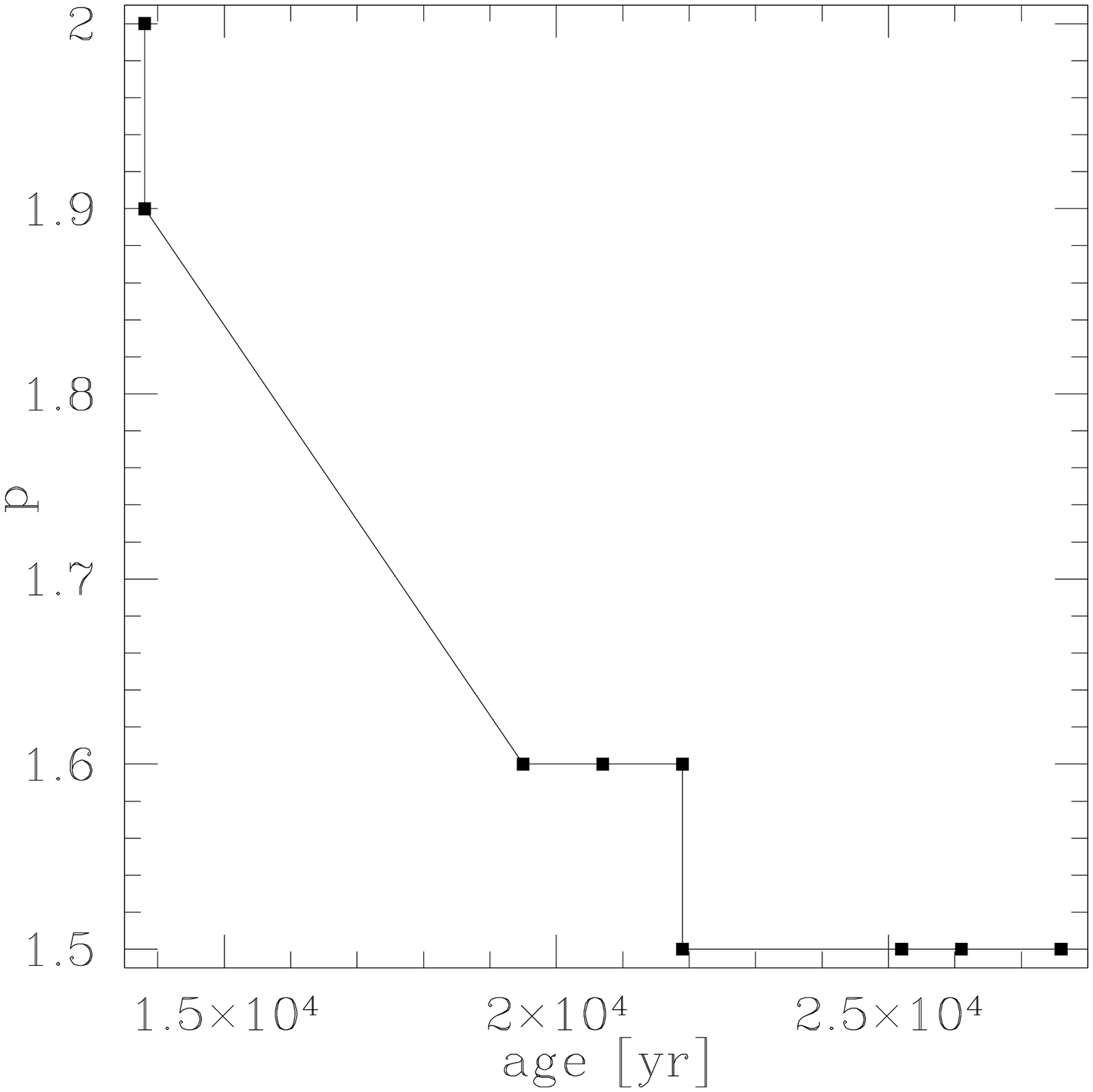}}
\caption{\label{mod2} Simulated and observed radial profiles as a function of
the impact parameter b(AU)  (left, short dashed and solid lines, respectively). The
dotted line represents the 1\,$\sigma$ noise level at the edge of the map.
Power-law index as function of time (right).}
\end{figure}

Detailed physical interpretation of the internal structure of nine objects was
carried out with intensive computer modeling. A simple spherically symmetric
model envelope, and assumptions about density and dust distributions following
the standard envelope model reproduce reasonably well the observed SED and the
radial profiles of the sources. $T \propto r^{-0.4}$ is a good approximation
for the sample. The radial temperature distribution as function of distance,
however, departs significantly from the optically-thin assumption, an
observational derivation of a single-power law q of 0.4 for radii $<$ 10 AU.
Modelling results indicate a density profile well described by a power-law
between p=1.5 to 2, which is expected by all of the collapse models and
numerical studies. A density structure of $\rho \propto r^{-2}$ at younger
ages, evolving to $\rho \propto r^{-3/2}$ at later times was found.

SED and radial profile fits can constrain physical parameters of the sources
such as envelope masses, density distributions, sizes and sublimation radii.
Nevertheless, observations at 10-300 $\mu$m, and the inclusion of outflow and
disk, and other geometries will decrease the differences.

 %
 %
 %
 %

\begin{center}
\scriptsize
\renewcommand{\tabcolsep}{4pt}
\begin{tabular}{lccccccccc}
\multicolumn{10}{c}{\normalsize\parbox{12cm}{Table\,1:
Best envelope fit results for the embedded sources. The temperature of the
central object T$_{*}$ is set to 3500\,K. $\chi^{2}$ quantifies the
agreement between model and data. The age is estimated according to 
the model of Smith (2000) and given in 10$^3$\,yrs.}} \\
\noalign{\smallskip} \hline \hline \noalign{\smallskip} 
Object & T$_{bol}$ [K] & L$_*$ [L$_\odot$] &
R$_{sub}$ [AU] & p & R$_{out}$ [AU]
& age & M$_{env}$\,[M$_{\odot}$] & $\chi^2$  \\
\noalign{\smallskip}
\hline \noalign{\smallskip}
L\,1448\,NW              &27    &~6 &5        &2.0            &~4000  &13.8 &~2.8  &~6.4 \\
L\,1448\,C               &40    &11 &5        &1.6            &~3000  &19.5 &~1.7  &~2.6 \\
RNO\,15\,FIR             &45    &~8 &3        &1.6            &~3500  &20.7 &~1.0  &~1.8 \\
NGC\,1333\,iras\,1       &52    &13 &3        &1.5            &~4500  &21.9 &~1.5  &~8.6 \\
NGC\,1333\,iras\,2       &48    &68 &3        &1.5            &~7000  &26.1 &~3.0  &~0.1 \\
HH\,211-mm               &30    &~5 &3        &1.9            &~6000  &13.8 &~4.1  &~4.5 \\
L\,1634                  &42    &19 &3        &1.6            &~6000  &21.9 &~2.8  &~1.1 \\
L\,1641\,N               &45    &85 &3        &1.5            &~8500  &25.2 &~7.0  &~1.0 \\
L\,1641\,sms\,III        &50    &80 &3        &1.5            &10000  &27.6 &~6.3  &~0.6 \\
\noalign{\smallskip} \hline \noalign{\smallskip}
\end{tabular}
\end{center}

 %
 %

\goodbreak

\References
\noindent Andr\'e, P., Ward-Thompson, D., Barsony, M. 1993, ApJ, 406, 122\\
Adams, F. 1991, ApJ, 382, 544\\
Froebrich, D. et al. 2003, MNRAS, in press\\
Rengel, M., Eisl\"{o}ffel, J., Hodapp, K.W. 2004a, in prep\\
Rengel, M., Froebrich, D., Wolf, S., Eisl\"{o}ffel, J. 2004b, in prep\\
Shirley, Y., Evans, N.J., Rawlings, J. 2002, ApJ, 575, 337\\
Smith, M.D. 2000, IAJ, 27, 1\\
Wolf, S., Henning, T., Stecklum, B. 1999, A\&A, 349, 839\\
\end{document}